\documentclass[11pt]{article}
\usepackage{cite}
\usepackage{amsmath, amsthm, amssymb,slashed}
\usepackage{ifpdf}
\ifpdf
  \usepackage[pdftex]{graphicx}
  \usepackage{epstopdf}
\else
  \usepackage[dvips]{graphicx}
\fi
\parskip=\baselineskip

\parskip=\baselineskip

\def\hat{\widehat}

\begin{document}

\begin{titlepage}

\vskip 1.5in

\begin{center}

{\bf\Large{On the Nature of Black Holes in Loop Quantum Gravity}}

\vskip0.5cm 

{Christian R\"oken\let\thefootnote\relax\footnotetext{e-mail: christian.roeken@cpt.univ-mrs.fr}} \vskip 0.05in {\small{ \textit{Centre de Physique Th\'{e}orique de Luminy}\vskip -.4cm
{\textit{Case 907, 13288 Marseille, France}}}

\vskip.05in{and}

\vskip .05in {\textit{Institut f\"ur Theoretische Physik, Lehrstuhl IV:\vskip -.4cm Weltraum- und Astrophysik, Ruhr-Universit\"at Bochum}}
\vskip-.15in {\textit{44780 Bochum, Germany}}}

\vskip.1in

{\small October 2012}

\end{center}

\vskip 0.5in
\baselineskip 16pt

\begin{abstract}

\noindent A genuine notion of black holes can only be obtained in the fundamental framework of quantum gravity resolving the curvature singularities and giving an account of the statistical mechanical, microscopic degrees of freedom able to explain the black hole thermodynamical properties. As for all quantum systems, a quantum realization of black holes requires an operator algebra of the fundamental observables of the theory which is introduced in this study based on aspects of loop quantum gravity. From the eigenvalue spectra of the quantum operators for the black hole area, charge and angular momentum, it is demonstrated that a strict bound on the extensive parameters, different from the relation arising in classical general relativity, holds, implying that the extremal black hole state can neither be measured nor can its existence be proven. This is, as turns out, a result of the specific form of the chosen angular momentum operator and the corresponding eigenvalue spectrum, or rather the quantum measurement process of angular momentum. Quantum mechanical considerations and the lowest, non-zero eigenvalue of the loop quantum gravity black hole mass spectrum indicate, on the one hand, a physical Planck scale cutoff of the Hawking temperature law and, on the other hand, give upper and lower bounds on the numerical value of the Immirzi parameter. This analysis provides an approximative description of the behavior and the nature of quantum black holes.

\end{abstract}

\end{titlepage}

\section{Introduction}

In loop quantum gravity, a recent attempt to reconcile the theories of general relativity and quantum mechanics into a single, consistent picture, black holes are usually described in the quasi-local frameworks of isolated and dynamical horizons (Ashtekar et al., 1998; Ashtekar, Baez \& Krasnov, 2000). Microscopically they are regarded as gases of non-interacting, distinguishable particles or rather topological boundary defects (Rovelli, 1996) with a discrete energy spectrum given by the area spectrum. The defects are caused by the polymer-like spin network excitations in the spacetime bulk structure that puncture the black hole horizon. 

\noindent These particular versions of quantum black holes have their origin in the spin network representation of loop quantum gravity (Rovelli \& Smolin, 1995). There, the quantum-gravitational states are described in terms of diffeomorphic equivalence classes of abstract graphs colored with irreducible representations of SU(2) on their links (spins) and invariant SU(2) intertwiners on their vertices. Conceptually, the graphs replace space on a fundamental level. One can think of spin networks as duals of cellular decompositions of space, where a certain volume is associated to a vertex and each boundary area with certain links. Given a black hole horizon with the topology $\Delta \simeq S^2 \times \mathbb{R}$, then its geometry is completely determined by the intersections of the graphs with its boundary (assuming that degenerate cases, where vertices of the graphs will be on the surface, do not exist). Labeling such intersections with $p \in \mathbb{N}^+ = \left\{1, 2, 3, ...\right\}$ and assigning to each link going through $p$ the color $j_p \in \mathbb{N}/2 = \left\{0, 1/2, 1, 3/2, ...\right\}$, one characterizes the quantum geometry of the surface by a $p$-tuple of spins $\boldsymbol{j} = (j_1, ..., j_p)$ giving rise to a total black hole surface area $A = 8 \pi l_p^2 \gamma \sum_{p} \sqrt{j_p (j_p + 1)}$, where $\gamma$ denotes the free parameter of loop quantum gravity, called Immirzi parameter, and $l_p = \sqrt{\hbar}$ is the Planck length in geometrical units $G = c = 1$. 

\noindent The thermal properties of black holes and the corresponding laws of black hole thermodynamics (Bardeen, Carter \& Hawking, 1973), discovered by Hawking and Bekenstein (Hawking, 1974 \& 1975; Bekenstein, 1973) studying quantum field theory in curved spacetimes in which effects coming from quantum geometry are not considered, could be recovered from a statistical mechanical account of the microscopic horizon degrees of freedom (Barreira, Carfora \& Rovelli, 1996; Krasnov, 1999a; Meissner, 2004; Agullo et al., 2010; Engle et al., 2010; Frodden, Ghosh \& Perez, 2011) in a quasi-local version in the framework of loop quantum gravity. This approach also discloses a new quantum hair (Ghosh \& Perez, 2011) as a purely quantum-geometrical phenomenon. The statistical mechanical analysis of isolated horizons fully complies with Hawking's and Bekenstein's findings since the quantum hair is irrelevant at the semiclassical level. 

\noindent In order to have a proper quantum description of black holes, one requires, as for all quantum systems, an operator algebra of fundamental observables. This study approaches this problem in the setting of loop quantum gravity following Bekenstein (2002) and provides such an algebra from a heuristic point of view, also clarifying the meaning and choice of the fundamental observables. As elements of this specific algebra, the black hole charge and angular momentum operators, that are used here, possess discrete sets of eigenvalues which, in conjunction with the eigenvalues of the area operator, lead to a quantum-gravitational modification of the upper bound on the extensive black hole parameter relation of classical general relativity. This suggests that it is impossible to measure an extremal black hole state with the choice of observables made. A discussion of this result and a comparison with another potential approach in the framework of loop quantum gravity (Bojowald, 2000), with a different angular momentum spectrum and, as a consequence, a different bound on the extensive parameters, is presented accordingly. Further, having the kinematical loop quantum gravity area spectrum available, a discrete mass spectrum for quantum black holes is derived. Including quantum mechanical considerations related to Heisenberg uncertainty principles, one is able to infer upper and lower bounds on the lowest, physical, non-zero black hole mass eigenvalue. This yields, at the same time, a restriction of the allowed values of the Immirzi parameter $\gamma$. As an implication of the lowest, physical black hole mass state in the final phase of the evaporation process, a Planck scale cutoff of the Hawking temperature law is in effect, which in turn constrains the values of the kinematical horizon area spectrum. Thus, in this paper some properties of quantum black holes are addressed within the loop formulation of quantum gravity leading to approximate solutions of several major issues in fundamental physics.

\section{Quantum Black Hole Algebra}

In classical general relativistic physics a generic, stationary black hole state is given by the Kerr-Newman vacuum solution of Einstein-Maxwell theory (Newman et al., 1965). The extensive black hole parameters, mass $M$, electric charge $Q$ and angular momentum $J$, are defined through the asymptotic behavior of the geometry and the electromagnetic field seen by an observer stationed at infinity. Disregarding more exotic charges like color, skyrmion number or the quantum numbers from non-abelian Yang-Mills or Proca fields, the no-hair conjecture postulates that all black hole solutions are completely determined by only these three classical quantities (Misner, Thorne, \& Wheeler, 1973) implying the loss of all information on the details of the inner structure, the topology and dynamical aspects. This indicates that the black hole interior is causally disconnected from its exterior. The coupling between both regions takes place only at the horizon. Therefore, it makes sense to consider the area of the event horizon boundary expressed as a function of $M, Q$ and $J$   

\begin{equation}\label{AREALAWBH}
A = 4 \pi \Bigl(r_+^2 + a^2\Bigr) = 8 \pi \Biggl(M \Biggl[M + \sqrt{M^2 - Q^2 - \frac{J^2}{M^2}} \,\, \Biggr] - \frac{Q^2}{2}\Biggr),
\end{equation}

\noindent where $r_+ = M + \sqrt{M^2 - Q^2 - J^2/M^2}$ denotes the Schwarzschild radius and $a = J/M$ the
angular momentum per unit black hole mass, as another classical observable of the Kerr-Newman black hole. This concurs with the view that an outside observer who is completely cut off from the inside region is just able to quantify the event horizon geometry as a practical black hole feature. In order to establish a proper algebraic quantum description for the dynamics of black holes, one has to identify the fundamental observables of the theory. The classical sector provides a set of four reasonable, equivalent parameters $\{M, Q, J, A\}$, with merely three of them independent. This raises the question of the choice of the trivalent subset of classical variables that is promoted to quantum operators functioning as fundamental observables of the quantum theory leaving the remaining variable corresponding to a secondary observable.

\subsection{Fundamental Observables}

The most general type of black hole is the charged and rotating Kerr-Newman black hole. Special cases are given by the rotating Kerr, the charged Reissner-Nordstr\"om and the static Schwarzschild black holes. They all have one non-zero feature in common, the mass $M$. One can have a black hole with $Q = J = 0$ and $M > 0$ but never with $M = 0$ and $|Q| > 0$ and/or $J > 0$. Therefore, charge and angular momentum are additional structures that can be imposed and are preferably fundamental observables in a quantum theory. One last question remains: Which of the two black hole parameters, mass $M$ or area $A$, should be selected for the role of the third fundamental variable? Given that from Eq.(\ref{AREALAWBH}) it is evident that there is, a priori, no difference in either choosing the mass or the area as quantum observable, the answer is simple: The choice depends, in an experimental setting, on the quantity one intends to measure as well as on the measurement device itself and, in a theoretical setting, on the available mathematical quantities (dynamical equations, spectra, ...) and their computability. Measuring or having a direct, theoretical eigenvalue spectrum of $M$ given by an ordered n-tuple $(M_1, ..., M_n)$, the specific information on the eigenvalues of area, charge and angular momentum, according to the $M$-inverse of Eq.(\ref{AREALAWBH}), enter only implicitly in the $M_i$-values. Thus, this $M$-spectrum encodes this detailed information. Then again, if one were to measure or have a direct, theoretical area spectrum at hand with values $(A_1, ..., A_n)$, this would encode the information on the mass $(M_1, ..., M_n)$, charge $(Q_1, ..., Q_n)$ and angular momentum $(J_1, ..., J_n)$ eigenvalues. Inverting Eq.(\ref{AREALAWBH}) with respect to $M$ would then yield the values $M_i(Q_i, J_i, A_i) = M_i(M_S, M_{RN}, M_K)$ that reveal something about the inner structure of the Kerr-Newman mass with Schwarzschild (S), Reissner-Nordstr\"om (RN) and Kerr (K) contributions, while the $A_i$ hide them. The same argument also applies for $A_i(M_i, Q_i, J_i)$ in the case of measuring or having a direct, theoretical mass spectrum. Both $(A_1(M_i, Q_i, J_i), ..., A_n(M_i, Q_i, J_i))$ (or $(M_1, ..., M_n)$) of the former and $(A_1, ..., A_n)$ (or $(M_1(Q_i, J_i, A_i), ..., M_n(Q_i, J_i, A_i))$) of the latter case are complete, equivalent descriptions of the area (or mass) spectrum, however, the first is indirect with respect to $A$ and direct with respect to $M$, while the second one is direct with respect to $A$ and indirect with respect to $M$. The central point of the analysis presented here, is the discrete, kinematical loop quantum gravity area spectrum accounting for the black hole event horizon surface in a simple, direct eigenvalue form $(A_1, ..., A_n)$ implicitly containing all information on the black hole mass, charge and angular momentum. Therefore, it is most suitable to take the area $A$ as the third fundamental observable.

\subsection{Construction of the Quantum Black Hole Algebra}

In the algebraic approach to black hole quantization introduced in this paper, one considers the operator of the black hole horizon area $\hat{A}_H$ alongside the operators of the electric charge $\hat{Q}$ and the angular momentum $\hat{J}$ as the fundamental quantum observables of black holes with elementary inputs and assumptions coming from loop quantum gravity and the requirements of local SO(3) as well as global U(1) gauge symmetries. The area operator $\hat{A}$ of a surface $S$ is the simplest realization of a geometric observable in loop quantum gravity (De Pietri \& Rovelli, 1996; Don\'a \& Speziale, 2010). It yields

\begin{equation}\label{LQGAS}
\hat{A}(S) \, \psi_{\Gamma} = 8 \pi l_p^2 \gamma \sum_{p \in S \cup \Gamma} \sqrt{j_p (j_p + 1)} \,\, \psi_{\Gamma}
\end{equation}

\noindent acting on spin network states $\psi_{\Gamma}$, with a discrete eigenvalue spectrum accounting for the area of $S$. The sum includes all spin contributions $j_p \in \mathbb{N}/2$ associated to the finite number of punctures $p$ caused by the intersections of the links of the spin network graph $\Gamma$ through $S$. The spin network states diagonalize the area operator and, thus, are eigenstates thereof. The lowest, non-vanishing area eigenvalue is the Planck scale or area gap $A_{min} = 4 \sqrt{3} \pi l_p^2 \gamma$ providing a physical ultraviolet cutoff for the surface degrees of freedom. Note that the geometrically motivated area spectrum (\ref{LQGAS}) resembles the square root of the angular momentum spectrum of the Casimir invariant $\hat{J}^{\, 2}$ of the three-dimensional rotation group suggesting that an internal Casimir operator $\hat{J}^{\, 2}_p$ can measure the quantum area of a given surface structure. This relation is discussed in detail in Krasnov \& Rovelli (2009) and in Bianchi (2011). They consider a system of $n$ particles each having a spin $j_p$ with states in a single-particle tensor product Hilbert space $\mathcal{H} = \mathcal{H}^{(j_1)} \otimes ... \otimes \mathcal{H}^{(j_n)}$. Simultaneous eigenstates of the $i$th component $\hat{J}^{\, i}_p$ of the angular momentum operator $\hat{J}_p$ and of the Casimir operator $\hat{J}^{\, 2}_p$ constitute an orthonormal basis for $\mathcal{H}$. These states are the aforementioned spin network states $\psi_{\Gamma}$. Since $\hat{J}^{\, i}_p$ and $\hat{J}^{\, 2}_p$ have eigenvalues of the forms $\sim m_p$ and $\sim j_p (j_p + 1)$, respectively, where $m_p$ is the spin projection quantum number of the $p$th link, which can, in general, take on the values $\left\{- j_p, - j_p + 1, ..., j_p - 1, j_p\right\}$, it makes sense to display the spin network states in the more explicit notation $\psi_{\Gamma} = \bigl|\{j_p, m_p\}_1^n; ... \bigr\rangle$ with $n = p_{max}$. It is important to stress that now both the spin $j_p$ and the spin projection quantum number $m_p$ label the links puncturing the surfaces of interest. Normally, the angular momentum of a black hole is related to SO(3) spatial symmetries and operations whereas the SU(2) spin network labels of the quantum-gravitational states refer to the internal rotation subgroup of the local Lorentz group. Hence, accounting for angular momentum in terms of an internal gauge group, as is done below, would, at first sight, seem odd since these concepts are not, a priori, linked to each other. If, however, the internal gauge group coincides with the group SU(2), which is the case in the connection variable formulation of general relativity typically used for loop quantum gravity, it can be related to angular momentum operations because the phase space structure at the black hole horizon boundary ties spatial rotations to internal dreibein rotations (Bojowald, 2000). Therefore, in this specific context, one has a correlation between the angular momentum of a black hole and the internal spin network labels at the horizon which allows for an angular momentum spectrum $J$ of a black hole, depending on the spin projection quantum numbers $m_p$ in the form   

\begin{equation}\label{LQGAMES}
J = l_p^2 \gamma \sum_{p = 1}^n m_p. 
\end{equation}

\noindent This spectrum can be inferred from the following semiclassical argument which is also used to set an upper bound on $J$ in terms of the spin labels $j_p$. A comparison of the angular momentum eigenvalues (\ref{LQGAMES}) with a different spectrum found in Bojowald (2000) and a discussion of the implications is presented at the end of Section 3.1.

\noindent Rotationally invariant spin network states fulfill the operator constraint 

\begin{equation}\label{JOC}
\sum_{p = 1}^n \hat{J}_p^{\, i} \bigl|\{j_p, m_p\}_1^n; ... \bigr\rangle = 0.
\end{equation}

\noindent The semiclassical limit of this equation becomes the closure condition 

\begin{equation}
\sum_{p = 1}^n j_p \boldsymbol{n}_p = 0,
\end{equation}
 
\noindent where $\boldsymbol{\hat{J}}_p \rightarrow \boldsymbol{j}_p \in \mathbb{R}^3$ and $j_p := ||\boldsymbol{j}_p||$ are the SU(2) spin labels. The vector $\boldsymbol{n}_p$ is a normal to the surface pierced by the $p$th link. It follows that 

\begin{equation}
\sum_{p = 1}^n j_p \geq 0.
\end{equation}

\noindent For states carrying angular momentum, one can deduce a generalization of Eq.(\ref{JOC}) for rotations around an internal direction constituting the axis of symmetry, for instance the $1$-direction, reading

\begin{equation}\label{ANGMOMEQ}
\sum_{p = 1}^n \hat{J}_p^{\, i} \bigl|\{j_p, m_p\}_1^n; ... \bigr\rangle = \frac{J}{l_p^2 \gamma} \, \delta^i_1 \bigl|\{j_p, m_p\}_1^n; ... \bigr\rangle, 
\end{equation}

\noindent where $J$ is given by (\ref{LQGAMES}). Then, an upper bound on the angular momentum eigenvalue spectrum of a rotating black hole can be obtained

\begin{equation}\label{AMB}
J \leq l_p^2 \gamma \sum_p j_p.
\end{equation}

\noindent This relation differs from the intuition of Krasnov (1999b) merely by the rescaling factor $\gamma$ which has to be introduced in order to preserve the angular momentum-area inequality for consistency with classical general relativity (see Section 3.1). With this loop quantum gravity setup in mind a black hole horizon area operator $\hat{A}_H$ as well as a black hole angular momentum operator $\hat{J}$ with actions

\begin{equation}\label{BHA}
\hat{A}_H \bigl|\{j_p, m_p\}_1^n; ... \bigr\rangle = A_H \bigl|\{j_p, m_p\}_1^n; ... \bigr\rangle = 8 \pi l_p^2  \gamma \sum_{p = 1}^n \sqrt{j_p (j_p + 1)} \bigl|\{j_p, m_p\}_1^n; ... \bigr\rangle 
\end{equation}

\noindent and

\begin{equation}\label{BHJ}
\sum_{p = 1}^n \hat{J}_p^{\, i} \bigl|\{j_p, m_p\}_1^n; ... \bigr\rangle = \frac{J}{l_p^2 \gamma} \, \delta^i_1 \bigl|\{j_p, m_p\}_1^n; ... \bigr\rangle = \delta^i_1 \sum_{p = 1}^n m_p \bigl|\{j_p, m_p\}_1^n; ... \bigr\rangle
\end{equation}

\noindent on the space of spin network states, where the latter satisfies the angular momentum bound (\ref{AMB}), can be defined. Together, they establish the quantum geometry of the black hole horizon. What follows in this subsection is done on the basis of an algebraic study by Bekenstein (2002) specifically carried out for black holes in loop quantum gravity. Further, the differences between Bekenstein's and this analysis are discussed.

\noindent A generic quantum black hole state $\bigl|BH\bigr\rangle$ has to be identified, in addition to the geometric numbers $j_p$ for the area and $m_p$ for the angular momentum, with a quantum number $q$ for the electric charge of the black hole ($e q$ is an eigenvalue of the charge operator $\hat{Q}$ with $q \in \mathbb{Z}$ and the elementary charge $e$) and a parameter $d \in \mathbb{N}^+$ determining the degree of degeneracy so that $\bigl|BH\bigr\rangle := \bigl|j_p, m_p, q, d\bigr\rangle$. These quantum numbers are necessary for a microscopic account of the distinguishable, internal states. One can define a vacuum black hole state by $\bigl|vac\bigr\rangle := \bigl|0, 0, 0, d\bigr\rangle$. Each state $\bigl|j_p, m_p, q, d\bigr\rangle$ correlates with an operator $\hat{\mathcal{Z}}_{j_p, m_p, q, d}$ such that 

\begin{equation}\label{PBHS}
\left|j_p, m_p, q, d\right\rangle = \hat{\mathcal{Z}}_{j_p, m_p, q, d} \left|vac\right\rangle.
\end{equation} 

\noindent Given the kinematical black hole configuration space spanned by the states $\left|j_p, m_p, q, d\right\rangle$, it appears sufficient to consider the operators $\hat{A}_H, \hat{Q}, \hat{J}, \hat{\mathcal{Z}}_{j_p, m_p, q, d}$ and the unity or identity operator $\hat{I}$ for a heuristic construction of an algebraic quantum theory of black holes. The foundation of this algebra is the set of operator equations 

\begin{equation}\label{OEFO}
\begin{split}
& \hat{A}_H \left|j_p, m_p, q, d\right\rangle = A_H \left|j_p, m_p, q, d\right\rangle \\ \\
& l_p \sqrt{\gamma} \hat{Q} \left|j_p, m_p, q, d\right\rangle = Q \left|j_p, m_p, q, d\right\rangle \\ \\
& l_p^2 \gamma \hat{J} \left|j_p, m_p, q, d\right\rangle = J \left|j_p, m_p, q, d\right\rangle \\ \\
& \hat{I} \left|j_p, m_p, q, d\right\rangle = \left|j_p, m_p, q, d\right\rangle,
\end{split}
\end{equation}

\noindent with the discrete spectra $A_H = 8 \pi l_p^2 \gamma \sum_{p = 1}^n \sqrt{j_p (j_p + 1)}$, $Q = l_p \sqrt{\gamma} e q$ and $J = l_p^2 \gamma \sum_{p = 1}^n m_p$, which is partially based on (\ref{BHA}) and (\ref{BHJ}), and the state operator $\hat{\mathcal{Z}}_{j_p, m_p, q, d}$. By definition of $\left|vac\right\rangle$, direct actions of $\hat{A}_H$, $\hat{Q}$ and $\hat{J}$ on the vacuum state lead to

\begin{equation}
\hat{A}_H \left|vac\right\rangle = \hat{Q} \left|vac\right\rangle = \hat{J} \left|vac\right\rangle = 0.
\end{equation}

\noindent Moreover, closure and linearity are imposed on the algebra, i.e., the commutators of any two operators of the algebra, on the one hand, do not produce new operators that are not already elements of the algebra and, on the other hand, are linear combinations of the operators inherently present. It is physically justified to demand that both area and charge should be invariant under SO(3) rotations and that the area should also be U(1) gauge invariant. Since the angular momentum is a measure for rotation and the electric charge is the generator of the U(1) symmetry of electromagnetism, i.e., of global gauge transformations, these requirements entail the commutators

\begin{equation}\label{BZC}
\bigl[\hat{A}_H, \hat{J} \, \bigr] = \bigl[\hat{Q}, \hat{J} \, \bigr] = \bigl[\hat{A}_H, \hat{Q} \, \bigr] = 0.
\end{equation}
 
\noindent From another perspective, because the area operator $\hat{A}_H$ is associated to the Casimir operator $\hat{J}^{\,2}$ of the angular momentum operator $\hat{J}$, which have simultaneous spin network eigenstates, it is obvious that $\hat{A}_H$ and $\hat{J}$ commute. Eq.(\ref{BZC}) is necessary for quantum black holes to have simultaneous area, charge and angular momentum eigenstates. This is required for having a full account on all relevant black hole characteristics. With $\hat{Q} \left|j_p, m_p, q, d\right\rangle = e q \, \left|j_p, m_p, q, d\right\rangle$, one can deduce a more general operator equation

\begin{equation}
e^{i \xi \hat{Q}} \hat{\mathcal{Z}}_{j_p, m_p, q, d} \left|vac\right\rangle = e^{i \xi e q} \hat{\mathcal{Z}}_{j_p, m_p, q, d} \left|vac\right\rangle,
\end{equation}

\noindent with $\xi \in \mathbb{R}$, which in turn gives, up to an approximation error of the order $\mathcal{O}(\xi)$, the basic commutator

\begin{equation}
\bigl[\hat{Q}, \hat{\mathcal{Z}}_{j_p, m_p, q, d}\bigr] = e q \hat{\mathcal{Z}}_{j_p, m_p, q, d}.
\end{equation}

\noindent All non-vacuum black hole states $\left|j_p, m_p, q, d\right\rangle$ must transform under SO(3) rotations as the corresponding spinorial harmonics $\boldsymbol{Y}$ and, hence, $\hat{\mathcal{Z}}_{j_p, m_p, q, d}$ must behave like an irreducible spinorial tensor operator of rank $j_p$ 

\begin{equation}
\bigl[\hat{J}, \hat{\mathcal{Z}}_{j_p, m_p, q, d}\bigr] = \Biggl[\sum_{p = 1}^n m_p\Biggr] \, \hat{\mathcal{Z}}_{j_p, m_p, q, d}.
\end{equation}

\noindent The horizon area operator $\hat{A}_H$ commutes with the black hole state operator $\hat{\mathcal{Z}}_{j_p, m_p, q, d}$ as follows 

\begin{equation}\label{AZ}
\bigl[\hat{A}_H, \hat{\mathcal{Z}}_{j_p, m_p, q, d}\bigr] = 8 \pi l_p^2 \gamma \sum_{p = 1}^n \sqrt{j_p (j_p + 1)} \, \hat{\mathcal{Z}}_{j_p, m_p, q, d}. 
\end{equation}

\noindent This can be shown with a straightforward calculation by the action of $\hat{A}_H$ on the vacuum state. For clarity it is useful to introduce the compact notation $\hat{\mathcal{Z}}_{\mu} := \hat{\mathcal{Z}}_{(j_{p})_{\mu}, (m_{p})_{\mu}, q_{\mu}, d_{\mu}}$ for a black hole state in the $\mu$-configuration. Then, according to the closure and linearity requirements, one obtains for the commutator of two black hole state operators

\begin{equation}\label{ZZ}
\bigl[\hat{\mathcal{Z}}_{\mu}, \hat{\mathcal{Z}}_{\nu}\bigr] = \sum_{\tau} a^{\tau}_{\,\,\,\mu \nu} \hat{\mathcal{Z}}_{\tau} + b_{\mu \nu} \hat{I} + c_{\mu \nu} \hat{A} + d_{\mu \nu} \hat{Q} + e_{\mu \nu} \hat{J},
\end{equation}

\noindent where all constants $a^{\tau}_{\,\,\,\mu \nu}, b_{\mu \nu}, c_{\mu \nu}, d_{\mu \nu}$ and $e_{\mu \nu}$ are antisymmetric in $\mu$ and $\nu$. As proposed in Bekenstein (2002), one assumes that for fixed $\mu \not= \nu$, $a^{\tau}_{\,\,\,\mu \nu} \not= 0$ for at least one index value $\tau$ and, therefore, $\hat{\mathcal{Z}}_{\mu}$ and $\hat{\mathcal{Z}}_{\nu}$ never commute. Commutativity would forbid scenarios where black hole mergers occur and, since it is known from the classical sector that they exist in nature, quantum non-commutativity is the simplest way of implementing this on a fundamental level. The Jacobi identity

\begin{equation}
\Bigl[\hat{A}_H, \bigl[\hat{\mathcal{Z}}_{\mu}, \hat{\mathcal{Z}}_{\nu}\bigr]\Bigr] + \Bigl[\hat{\mathcal{Z}}_{\mu}, \bigl[\hat{\mathcal{Z}}_{\nu}, \hat{A}_H\bigr]\Bigr] + \Bigl[\hat{\mathcal{Z}}_{\nu}, \bigl[\hat{A}_H, \hat{\mathcal{Z}}_{\mu} \bigr]\Bigr] = 0,
\end{equation}

\noindent together with Eqs.(\ref{AZ}) and (\ref{ZZ}), determines the constants $b_{\mu \nu} = c_{\mu \nu} = d_{\mu \nu} = e_{\mu \nu} = 0$ and results in

\begin{equation}\label{ZAR}
\hat{A}_H \bigl[\hat{\mathcal{Z}}_{\mu}, \hat{\mathcal{Z}}_{\nu}\bigr] = \sum_{\tau} a^{\tau}_{\,\,\, \mu \nu} \, \left[A_{H}\right]_{tot} \, \hat{\mathcal{Z}}_{\tau} = 8 \pi l_p^2 \gamma \Biggl(\Biggl[\sum_{p = 1}^n \sqrt{j_p (j_p + 1)}\Biggr]_{\mu} + \Biggl[\sum_{p = 1}^{n'} \sqrt{j_p (j_p + 1)}\Biggr]_{\nu}\Biggr) \bigl[\hat{\mathcal{Z}}_{\mu}, \hat{\mathcal{Z}}_{\nu}\bigr]
\end{equation}

\noindent with $\left[A_{H}\right]_{tot} = \left[A_{H}\right]_{\mu} + \left[A_{H}\right]_{\nu}$. The same procedure can be carried out for $\hat{Q}$ and $\hat{J}$ which yields 

\begin{equation}\label{ZCH}
\hat{Q} \bigl[\hat{\mathcal{Z}}_{\mu}, \hat{\mathcal{Z}}_{\nu}\bigr] = e \bigl(q_{\mu} + q_{\nu}\bigr) \bigl[\hat{\mathcal{Z}}_{\mu}, \hat{\mathcal{Z}}_{\nu}\bigr]
\end{equation}

\noindent and

\begin{equation}\label{ZAM}
\hat{J} \bigl[\hat{\mathcal{Z}}_{\mu}, \hat{\mathcal{Z}}_{\nu}\bigr] = \Biggl(\Biggl[\sum_{p = 1}^n m_{p}\Biggr]_{\mu} + \Biggl[\sum_{p = 1}^{n'} m_{p}\Biggr]_{\nu}\Biggr) \bigl[\hat{\mathcal{Z}}_{\mu}, \hat{\mathcal{Z}}_{\nu}\bigr].
\end{equation}

\noindent The hermitian conjugate of the black hole state operator $\hat{\mathcal{Z}}_{\mu}^{\dagger}$ has the characteristics of an annihilation operator 

\begin{equation}
\bigl[\hat{A}_H, \hat{\mathcal{Z}}_{\mu}^{\dagger}\bigr] = - \left[A_{H}\right]_{\mu} \, \hat{\mathcal{Z}}_{\mu}^{\dagger}.
\end{equation}

\noindent Because of this, it directly becomes obvious that the following three equations hold 

\begin{equation}\label{ZANH}
\begin{split}
& \hat{A}_H \bigl[\hat{\mathcal{Z}}_{\mu}^{\dagger}, \hat{\mathcal{Z}}_{\nu}\bigr] = 8 \pi l_p^2 \gamma \Biggl(\Biggl[\sum_{p = 1}^{n'} \sqrt{j_p (j_p + 1)}\Biggr]_{\nu} - \Biggl[\sum_{p = 1}^n \sqrt{j_p (j_p + 1)}\Biggr]_{\mu}\Biggr) \bigl[\hat{\mathcal{Z}}_{\mu}^{\dagger}, \hat{\mathcal{Z}}_{\nu}\bigr] \\ \\
& \hat{Q} \bigl[\hat{\mathcal{Z}}_{\mu}^{\dagger}, \hat{\mathcal{Z}}_{\nu}\bigr] = e \bigl(q_{\nu} - q_{\mu}\bigr) \bigl[\hat{\mathcal{Z}}_{\mu}^{\dagger}, \hat{\mathcal{Z}}_{\nu}\bigr] \\ \\
& \hat{J} \bigl[\hat{\mathcal{Z}}_{\mu}^{\dagger}, \hat{\mathcal{Z}}_{\nu}\bigr] = \Biggl(\Biggl[\sum_{p = 1}^{n'} m_{p}\Biggr]_{\nu} - \Biggl[\sum_{p = 1}^{n} m_{p}\Biggr]_{\mu}\Biggr) \bigl[\hat{\mathcal{Z}}_{\mu}^{\dagger}, \hat{\mathcal{Z}}_{\nu}\bigr].
\end{split}
\end{equation}

\noindent From Eqs.(\ref{ZAR}), (\ref{ZCH}), (\ref{ZAM}) and (\ref{ZANH}) it can be concluded that $\bigl[\hat{\mathcal{Z}}_{\mu}, \hat{\mathcal{Z}}_{\nu}\bigr]$ and $\bigl[\hat{\mathcal{Z}}_{\mu}^{\dagger}, \hat{\mathcal{Z}}_{\nu}\bigr]$ are physical black hole states as well. The quantum black hole algebraic structures given in Bekenstein (2002) and in the study at hand differ mainly in the specific forms of the area and the angular momentum spectra. Bekenstein makes the simple ad hoc assumption that the area is quantized in multiple integers of an elementary, Planck-sized area $A_0$ according to the formula $A_k = k A_0$, where $k \in \mathbb{N}$. This ansatz is reasonable because it represents the idea of quantum geometry with fundamental quanta of area. Nonetheless, it leads to a discrete black hole mass emission spectrum which is completely unlike Hawking's thermal emission spectrum (for more details see, e.g., Bekenstein \& Mukhanov (1995) and Barreira, Carfora \& Rovelli (1996)). With the loop quantum gravity area spectrum, one yields a quasi-continuous black hole mass spectrum that does not have this problem (see also Section 3.2). Thus, in the approach applied here, a well-defined area operator eigenvalue spectrum as a clear and sound result from loop quantum gravity was used and no assumptions and speculations regarding this spectrum were made. Besides, while in Bekenstein's analysis there is an ab initio postulated angular momentum algebra that is related to SO(3) spatial rotations without, of course, any mention of spin networks, here, a connection between the angular momentum of a black hole and rotations in an internal gauge group structure is exploited. The explicit knowledge of how Kerr-Newman quantum black hole states behave is essential because quantum dynamics is at the bottom of every physical system and it is, therefore, imperative to have a fundamental picture of the nature of black hole quantization. The algebraic approach established here from heuristic, spectral considerations and assumptions emerging from loop quantum gravity and specific symmetries of the fundamental, discrete observables, gives a basic perspective on this topic.

\section{Further Loop Heuristics}

\subsection{Extremal Black Holes and Bound on Extensive Black Hole Parameters}

\noindent The cosmic censorship conjecture (Wald, 1984) of classical general relativity roughly states that black holes, emerging from gravitational collapse, are always expected to be hidden behind a horizon and never to produce a naked singularity which is considered unphysical. This imposes the classical, physically motivated bound

\begin{equation}
M^2 \geq Q^2 + \frac{J^2}{M^2}
\end{equation}

\noindent which in terms of the black hole horizon area $A_H$, instead of the mass $M$, reads

\begin{equation}\label{AC}
A_H \geq 4\pi \sqrt{Q^4 + 4 J^2}.
\end{equation}

\noindent This bound gives a restriction to just those situations one anticipates to apply in real gravitational collapses. The equality holds for the case known as extremal black hole, where area, charge and angular momentum are balanced. This type of black hole appears to be quite unstable because even a very small perturbation can bring it to cross over to the forbidden region $A_H < 4\pi \sqrt{Q^4 + 4 J^2}$. The area spectrum in Eq.(\ref{BHA}) and the restriction (\ref{AMB}) on the angular momentum spectrum (\ref{BHJ}), however, indicate that black holes with extremal parameters can neither be measured nor can their existence be proven using the corresponding set of operators, thus, avoiding the issue of black hole instability. In order to see this, it is sufficient to limit the discussion to black holes of the Kerr family with $Q = 0$, which simplifies, but does not change the nature of the argument presented. The classical general relativistic constraint (\ref{AC}) then becomes 

\begin{equation}\label{SBHAB}
A_H \geq 8 \pi J.
\end{equation}  

\noindent From the discrete area spectrum $A_H$, stated on the right-hand side of Eq.(\ref{BHA}), and the angular momentum bound (\ref{AMB}), one can infer the strict inequality 

\begin{equation}\label{SE}
A_H = 8 \pi l_p^2 \gamma \sum_{p = 1}^n \sqrt{j_p (j_p + 1)} > 8 \pi l_p^2 \gamma \sum_{p = 1}^n j_p \geq 8 \pi J.
\end{equation}

\noindent This shows that the classical general relativistic bound (\ref{SBHAB}) for Kerr black holes can never be saturated. In the classical limit $j_p \rightarrow \infty$ on the other hand, where the area spectrum is of the form 

\begin{equation}
A_H^{Cl} = 8 \pi l_p^2 \gamma \sum_{p = 1}^n j_p, 
\end{equation}

\noindent one recovers inequality (\ref{SBHAB}), thus, making a saturation possible. But since the actual, physical area is given by the quantum spectrum $A_H$ in Eq.(\ref{BHA}), the existence of an extremal black hole state is impossible to detect with the particular set of quantum observables used. For the purpose of phenomenological studies, one can compute the first-order quantum-geometrical correction to inequality (\ref{SBHAB}) in order to obtain an effective representation of inequality (\ref{SE}). Based on the latter strict inequality, the leading-order loop quantum gravity correction $C_{LQG}$ to the classical horizon area
  
\begin{equation}
A_H^{SCl} = A_H^{Cl} + C_{LQG} > 8 \pi J
\end{equation}

\noindent reads

\begin{equation}
C_{LQG} = 2 \pi l_p^2 \gamma n (n + 1),
\end{equation}

\noindent where $n$ is the total number of punctures of the horizon. This formula is not valid at small scales (of the order of the Planck area $l_p^2$) anymore. It ignores quantum-geometrical substructures and short distance degrees of freedom yielding an approximation just including the appropriate semiclassical degrees of freedom. The derivation of a black hole angular momentum-area relation like inequality (\ref{SE}) depends strongly on the underlying angular momentum eigenvalue spectrum. Following a study by Bojowald of angular momentum in loop quantum gravity (Bojowald, 2000), one finds an angular momentum-area inequality that, unlike inequality (\ref{SE}), includes the extremal black hole case. This discrepancy is a direct consequence of the quantum measurement process as can be seen below. Bojowald introduces an angular momentum operator $\hat{L}$ in the framework of loop quantum gravity using a relation with spherically symmetric quantum-gravitational states, acting as non-rotating reference frames, for the definition of angular momentum. This operator can be diagonalized simultaneously with the area operator. Any of its components has the eigenvalues $L_i = \hbar \sum_p m_p$, while the spectrum of the absolute value in the spin-$j$ representation reads $L^{(j)} = \hbar \sqrt{j (j + 1)}$. The total angular momentum, for a given spin network state with spins $\{j_p\}$ describing the geometry of the black hole, is bounded from above $L \leq \hbar \sqrt{\sum_p j_p (\sum_p j_p + 1)}$. This inequality leads to $A_H \geq 8 \pi \gamma L$. Since the aforementioned spectra are not subjected to the same rescaling, by means of the Immirzi parameter $\gamma$, as those considered in the present analysis, this angular momentum-area inequality is not consistent with the relation from classical general relativity. The fact that the angular momentum-area bound of Bojowald's study allows for the occurrence of extremal black holes comes from choosing the total angular momentum or rather the square root of the Casimir spectrum $L$ for the values of the angular momentum of the quantum black hole. Applying an angular momentum operator $\hat{L}_{int}$ that corresponds to a measurement of the black hole angular momentum along an internal, normalized direction representing the axis of symmetry, as is done here, yields the strict inequality (\ref{SE}). In classical black hole physics, one can always construct a coordinate system in which the axis of rotation is collinear to one of the basis vectors and, thus, the classical observables $L_{int}$ and $L$ coincide. In a quantum theory, this is not true anymore. Measuring the angular momentum component along the axis of rotation creates uncertainties in the remaining components and, therefore, one obtains a different eigenvalue spectrum as in the case of detecting the total angular momentum. While in a classical theory these variables are the same, their quantum counterparts constitute different observables. Accordingly, an extremal black hole can neither be measured nor can its existence be verified or falsified, respectively, using the angular momentum component projected onto the axis of rotation. Theoretically, however, it can be seen with the help of a total angular momentum operator. The motivation to chose the former operator for the description of black hole rotations originates in the fact that its eigenvalues (\ref{LQGAMES}) define deformations and shifts in the area of the black hole horizon.

\subsection{Final Phase of the Black Hole Evaporation Process, Hawking Temperature Law Cutoff and Upper and Lower Bounds on the Immirzi Parameter $\gamma$}

Due to quantum mechanical effects near the event horizon, found in a semiclassical analysis examining quantum scalar field theory in curved spacetimes, black holes are supposed to emit a certain form of radiation, the Hawking radiation, which has, for sufficiently heavy black holes, the characteristic, continuous frequency spectrum of a black body (Hawking, 1974 \& 1975). This radiant emittance reduces the mass of the black hole and provides, therefore, a way for a black hole to evaporate. Bekenstein (1973) predicted that black holes should have finite, non-zero temperature and entropy. This radiation can be seen as a result of such thermodynamical properties of black holes. It is also required by the Unruh effect and the equivalence principle when applied to black hole horizons. The main component of this radiation is thermal. Since the temperature of a solar mass black hole is extremely low, its thermal energy is not high enough to exceed the rest energy of the lightest, massive particles so that effectively only thermal radiation (massless photons) can be produced. Nonetheless, when the black hole slowly evaporates due to this process ($t_{ev} \sim 10^{74} \, \textnormal{s}$ for a solar mass black hole) it becomes hotter according to the Hawking temperature law for static black holes

\begin{equation}\label{HTL}
T_H = \frac{\hbar}{8 \pi k_B \, M},
\end{equation}

\noindent thus, at a certain stage, the thermal energy of the black hole becomes sufficiently high in order for Hawking radiation to consist of various elementary particles. In the final phase of the black hole evaporation process, when the remaining black hole mass becomes extremely small, the temperature, according to Eq.(\ref{HTL}), increases rapidly and reaches unphysical values. In the following, a heuristic chain of arguments is used to show that a Planck scale cutoff of the black hole temperature is in effect. Considering the evaporation process of a quantum black hole, it must obey a Heisenberg uncertainty relation 

\begin{equation}\label{HUP}
\Delta E \Delta t \geq \frac{\hbar}{2}.
\end{equation} 

\noindent By interpreting the quantum uncertainties, as was already done by Nils Bohr in the famous Bohr-Einstein debates, as intervals over which specific physical quantities of the quantum object are spread and, therefore, giving a measure of their extension, $\Delta E$ determines the rest energy of the black hole and $\Delta t$ its complete evaporation time. It should be pointed out that the interpretation of an uncertainty as a measure of the total extension is only valid for the ultimate transition to the zero-eigenvalue of the quantity of interest. With the Hawking particle, which is emitted during the evaporation, having the proportion $\Delta x$ and carrying the momentum $\Delta p = \Delta E$, thus, taking away the residual quantum black hole mass leaving a state devoid of any black hole, one associates another Heisenberg uncertainty relation 

\begin{equation}\label{HUP2}
\Delta x \Delta p \geq \frac{\hbar}{2}.
\end{equation}

\noindent The applications of both uncertainty principles (\ref{HUP}) and (\ref{HUP2}) to an ultimate Hawking radiation burst yield upper and lower bounds on the minimal black hole mass. Substituting $\Delta M$ for the quantum black hole rest energy $\Delta E$ and $5120 \pi (\Delta M)^3/\hbar$ for the evaporation time $\Delta t$ in inequality (\ref{HUP}) leads to the lower minimal mass bound 

\begin{equation}\label{LB}
M_{min} = \Delta M \gtrsim 0.075 M_p,
\end{equation}

\noindent where $M_p$ denotes the Planck mass. With the Compton wavelength $\Delta x = h/\Delta M$ and the momentum $\Delta p = k_B \Delta T = \hbar/(8 \pi \Delta M)$ of the Hawking particle, one obtains, based on inequality (\ref{HUP2}), the upper minimal mass bound

\begin{equation}\label{UB}
M_{min} = \Delta M \lesssim 0.707 M_p.
\end{equation}

\noindent Since inequality (\ref{LB}) can be understood as a lower bound on the physical mass of a quantum black hole before it completely evaporates in a final emission carrying away a mass portion that is at least $M_{min} = 0.075 M_p$ (but bounded from above with a limit given by $0.707 M_p$), one cannot find any low-mass quantum black holes with minimal masses below $M = 0.075 M_p$. One can reach the same conclusion on a minimal black hole mass via order of magnitude estimates from loop quantum gravity. There, the smallest, physical length scale is of the order of the Planck length $l_p \sim 10^{-35}\,\textnormal{m}$ which has the consequence that, when the Schwarzschild radius of a black hole reaches this critical value, its mass will be of the order $M \sim M_p$. Since, according to this particular model of canonical quantum gravity, one cannot find a smaller, physical length scale in nature, a black hole cannot have a mass that falls below the Planck mass in accordance with the former quantum mechanical argument. It is, therefore, relevant and necessary for physical black hole applications to limit the values of the spin quantum numbers $j_p$ of the kinematical loop quantum gravity area spectrum from having elements in $\mathbb{N}/2$ to $\mathbb{N}^+/2$. A proper computation of the full dynamical horizon area spectrum is, hence, expected to show that the spin value $j_p = 0$ has to be excluded. Note that this restriction on the spin labels does not contradict the black hole vacuum state $\left|vac\right\rangle$ because this state represents merely a mathematical instrument for the construction of the physical black hole states (\ref{PBHS}). Nothing in the algebraic analysis suggests that $\left|vac\right\rangle$ is a physical black hole state. Accordingly, one has a lowest, physical black hole state with a minimal mass $M_{min} \sim M_p$ which suggests that an ultraviolet cutoff of the Hawking temperature at $M = M_{min}$ is required, ensuring a well-defined behavior of formula (\ref{HTL}). For $M = 0$, after the final Hawking emission did occur, the black hole ceases to exist and the Hawking temperature law loses its validity. One can take a more comprehensive look at the issue of the black hole mass levels from the point of view of loop quantum gravity heuristics. A mass spectrum $M$ can be computed by means of the Christodoulou-Ruffini formula (Christodoulou \& Ruffini, 1971), the $M$-inverse of Eq.(\ref{AREALAWBH}),

\begin{equation}
M = \sqrt{\frac{A}{16 \pi} \biggl(1 + \frac{4 \pi Q^2}{A} \biggr)^2 + \frac{4 \pi J^2}{A}}
\end{equation} 

\noindent and the discrete spectra of area $A = A_H$, charge $Q$ and angular momentum $J$ which are stated below Eq.(\ref{OEFO}), yielding 

\begin{equation}\label{FMASSBH}
M = l_p \sqrt{\frac{\gamma}{2 \sum_{p = 1}^n \sqrt{j_p (j_p + 1)}}} \times \sqrt{\Biggl(\sum_{p = 1}^n \sqrt{j_p (j_p + 1)} + \frac{e^2 q^2}{2} \Biggr)^2 + \Biggl(\sum_{p = 1}^n m_p \Biggr)^2}. 
\end{equation}

\noindent This gives reason for a secondary operator equation for a mass operator $\hat{M}$ of a quantum black hole

\begin{equation}
l_p \sqrt{\gamma} \hat{M} \left|j_p, m_p, q, d\right\rangle = M \left|j_p, m_p, q, d\right\rangle 
\end{equation}

\noindent in accordance with Eq.(\ref{OEFO}) for the fundamental observables $\{Q, J, A\}$. The mass spectrum of a Schwarzschild-type black hole is, in SI units ($l_p \rightarrow M_p$) and, thus, discretized in units of the Planck mass $M_p$,

\begin{equation}\label{MASSBH}
M(Q = J = 0) = M_p \sqrt{\frac{\gamma}{2} \sum_{p = 1}^n \sqrt{j_p (j_p + 1)}}.
\end{equation}

\noindent Then, the lowest, non-zero mass eigenvalue reads 

\begin{equation}\label{LMEV}
M_{min} = \frac{M_p \sqrt{\gamma \sqrt{3}}}{2}, 
\end{equation}

\noindent while higher eigenvalues also result in multiples of $M_p$ according to Eq.(\ref{MASSBH}). Only in the large $j_p$ limit confirms this spectrum Bekenstein's presumption of a mass spectrum of the form $M_k \sim \sqrt{\hbar k}$, with $k \in \mathbb{N}$, emerging from an equispaced area spectrum for quantum black holes. The transition from one black hole mass level to another can, in general, only be realized by the particular quantum mass jumps given by Eq.(\ref{MASSBH}). The average spacing between the mass levels decreases exponentially with $M$ (Barreira, Carfora \& Rovelli, 1996). Note that the conclusions drawn from Eqs.(\ref{LB}) and (\ref{UB}) are, therefore, only valid for the final transition $M_{min} \rightarrow 0$ in the quantum black hole regime. Comparing the lowest, non-zero eigenvalue (\ref{LMEV}) of the loop quantum gravity black hole mass spectrum (\ref{MASSBH}) to the lower and upper bounds (\ref{LB}) and (\ref{UB}), respectively, restricts the numerical value of the Immirzi parameter in the following way

\begin{equation}
0.013 \lesssim \gamma \lesssim 1.154.
\end{equation} 

\noindent This rough estimate includes all those values for $\gamma$ arising from studies of the thermal properties of isolated horizons ($\gamma \approx 0.127$ in Ashtekar et al. (1998); $\gamma \approx 0.238$ in Meissner (2004); $\gamma \approx 0.274$ in Agullo et al. (2010)), of quasinormal modes of black holes ($\gamma \approx 0.124$ in Dreyer (2003)) or of semiclassical, statistical mechanical descriptions in terms of shapes of tessellated horizons and conformations of closed polymer chains ($\gamma \approx 0.274$ in Bianchi (2011)). The Planck scale cutoff at $M_{min}$ obtained through heuristic considerations from quantum mechanics and loop quantum gravity, respectively, stops the black hole from getting infinite temperature in the final phase of the evaporation process. The black hole can only reach a finite, maximum temperature of $T_{max} = \hbar/(8 \pi k_B M_{min}) \sim T_p$. Thus, before the total evaporation, one is left, however, with a minimal black hole of extreme, but finite, physical characteristics of Planck-order.

\section{Discussion and Conclusions}

Loop quantum gravity provides a discrete, quantum-geometrical eigenvalue spectrum for an area operator $\hat{A}$ for arbitrary surfaces that can be used to characterize the quantum structure of black hole horizons. The horizon surface, as a fundamental observable in terms of this operator, can be used, along with the black hole charge and angular momentum operators $\hat{Q}$ and $\hat{J}$, for a heuristic, spectral study of quantum black holes giving an algebraic account of the simultaneous relations between these operators. Spectral estimates on the possible configurations of the area, charge and angular momentum eigenvalues as well as bounds on the lowest, physical black hole mass level yield, in particular, kinematical and dynamical aspects that shed new light on and leads to a more accurate description of the evolution of black holes. Following this heuristic sentiment, it could be shown that, by using a projection angular momentum operator as quantum observable, one cannot measure the extremal black hole state since the extensive parameters fulfill a strict cosmic censorship inequality. Moreover, black holes have a minimal, non-zero mass that is given by the lowest, physical mass eigenvalue derived from the kinematical loop quantum gravity area spectrum via the Christodoulou-Ruffini formula introducing a finite, ultraviolet cutoff of the Hawking temperature. This indicates that in the dynamical black hole horizon area spectrum at least the spin $j_p = 0$ has to be excluded. Quantum theoretical considerations relating to the minimal black hole mass regulate the free Immirzi parameter $\gamma$ of loop quantum gravity to have a numerical value approximately between $0.013$ and $1.154$. This interval contains all the values obtained in more sophisticated approaches. Accordingly, a coherent picture of the different stages in the life of a black hole emerges. After the gravitational collapse, infalling matter and radiation present in its vicinity serve as a "food source", while the black hole itself emits a steady flux of Hawking radiation. This mass-energy is emitted/absorbed only in discrete portions according to the mass spectrum (\ref{FMASSBH}), i.e., the black hole is only allowed to make quantized jumps in its mass levels. Since the density of the spectral lines grows exponentially with the black hole mass, one recovers, for large black holes, a quasi-continuous Planckian emission spectrum. When all the surrounding matter and radiation is finally absorbed, the black hole can begin its true evaporation. There exists no process able to suppress this. It starts to shrink at a rate that depends on its extensive parameters, first very slowly with purely thermal emissions, then faster and faster, while also non-zero rest mass particles can be produced. Before the total evaporation, it can reside in a minimal black hole state exhibiting purely Planck-order characteristics.

\vskip 0.9cm

\noindent {\it Acknowledgments}
 
\noindent I would like to thank Katharina Proksch, Horst Fichtner, Bastian Weinhorst and Michal Michno for many useful discussions and their comments. This research was supported by a DFG Research Fellowship.

\bibliographystyle{unsrt}

\end{document}